# Wilson Ratio of a Tomonaga-Luttinger Liquid in a Spin-1/2 Heisenberg Ladder

K. Ninios,[1,2] Tao Hong,[3] T. Manabe,[4] C. Hotta,[4] S. N. Herringer,[5]
M. M. Turnbull,[5] C. P. Landee,[5] Y. Takano,[1] and H. B. Chan[2]

[1]*Department of Physics, University of Florida, Gainesville, Florida 32611, USA*
[2]*Department of Physics, The Hong Kong University of Science and Technology, Clear Water Bay, Kowloon, Hong Kong, China*
[3]*Quantum Condensed Matter Division, Oak Ridge National Laboratory, Oak Ridge, Tennessee 37831, USA*
[4]*Department of Physics, Faculty of Science, Kyoto Sangyo University, Kyoto 603-8555, Japan*
[5]*Carlson School of Chemistry and Department of Physics,
Clark University, Worcester, Massachusetts 01610, USA*

(Dated: March 18, 2012)

Using micromechanical force magnetometry, we have measured the magnetization of the strong-leg spin-1/2 ladder compound $(C_7H_{10}N)_2CuBr_2$ at temperatures down to 45 mK. Low-temperature magnetic susceptibility as a function of field exhibits a maximum near the critical field $H_c$ at which the magnon gap vanishes, as expected for a gapped one-dimensional antiferromagnet. Above $H_c$ a clear minimum appears in the magnetization as a function of temperature as predicted by theory. In this field region, the susceptibility in conjunction with our specific heat data yields the Wilson ratio $R_W$. The result supports the relation $R_W = 4K$, where $K$ is the Tomonaga-Luttinger-liquid parameter.

PACS numbers: 75.10.Jm 71.10.Pm 75.78.-n 75.50.Ee

One of the crucial parameters that characterize Fermi liquids, such as conduction electrons and liquid $^3$He at low temperatures, is the Wilson ratio [1]

$$R_W = \frac{4}{3}\left(\frac{\pi k_B}{g\mu_B}\right)^2 \frac{\chi}{C/T}, \qquad (1)$$

the dimensionless ratio of the temperature-independent magnetic susceptibility $\chi$ to the coefficient of the $T$-linear specific heat $C$. By dividing out the contribution of enhanced mass, which enters both $\chi$ and $C$, this parameter quantifies spin fluctuations that enhance the susceptibility. For this reason, it serves as a powerful tool to classify heavy fermion systems [2]. For instance, $R_W$ is 2 for the $S=1/2$ Kondo lattice in the single-impurity limit [1] in contrast to 1 for non-interacting fermions.

In liquid $^3$He, the archetypical Fermi liquid, $R_W$ varies only weakly with pressure unlike the strongly pressure-dependent effective mass and is close to 4, the approximate limiting value for the Hubbard model with critical on-site repulsion [3]. This has promoted the view that quasiparticles in liquid $^3$He are nearly localized [4].

In one dimension (1D), where an arbitrarily weak interaction renders the quasiparticle lifetime shorter than $\hbar/(E-E_F)$ because of stringent constraints the spatial dimension imposes on energy and momentum conservation in scattering processes, the Fermi liquid theory breaks down entirely, giving way to the Tomonaga-Luttinger liquid (TLL) as the correct low-energy description of fermions [5]. In a TLL, low-lying excitations are massless, collective bosonic modes instead of fermionic quasiparticles endowed with effective mass. Nonetheless, the TLL and the Fermi liquid are alike in that in both systems, $\chi$ is independent of temperature $T$ and $C$ is linear in $T$. As a result, the Wilson ratio remains a crucial parameter in 1D—in fact even more crucial than in three dimensions, since a large variety of interacting systems in 1D fall into the TLL universality class and since each branch of bosonic modes in a given TLL is completely specified by just two parameters, the velocity $v$ and the TLL parameter $K$ [5, 6].

For instance, the Wilson ratio of 1D electrons contains only the velocities of spin and charge excitations, $v_\sigma$ and $v_\rho$, and the TLL parameter $K_\sigma = 1$ of the spin sector: $R_W = 2K_\sigma(1+v_\sigma/v_\rho)^{-1}$ [7]. In the non-interacting limit, where $v_\sigma = v_\rho = v_F$, $R_W$ is again 1 as expected. With on-site Coulomb repulsion, $v_\sigma$ decreases whereas $v_\rho$ increases, resulting in $R_W$ approaching 2 in the strongly repulsive limit [7, 8].

Another example of a TLL is a 1D antiferromagnet in a gapless regime—such as a spin-1 linear chain and a spin-1/2 ladder, both in magnetic fields larger than the magnon gap, or a spin-1/2 linear chain in zero field as well as in magnetic fields [6]. There, $\chi = (g\mu_B)^2 K/(\pi v)$ and $C = \pi k_B^2 T/(3v)$, and thus the Wilson ratio must obey the relation

$$R_W = 4K, \qquad (2)$$

which to our knowledge has not been noted before [9]. It is remarkable that except for the trivial numerical factor, the Wilson ratio is the TLL parameter $K$, which governs the exponents of all spin correlation functions, including dynamic ones. This is not surprising, however, since TLLs are quantum critical [10] and, consequently, their dynamic properties are inextricably linked to static properties [11].

Despite the significance illustrated by these examples, the Wilson ratio has never been determined experimentally in a TLL because of the lack of a good material. Here we report the Wilson ratio of a strong-leg $S=1/2$ Heisenberg spin-ladder antiferromagnet in magnetic fields, determined from accurate micromechanical force magnetometry and specific-heat measurements. The result is in agreement with a density-matrix renormalization-group calculation in conjunction with Eq. 2.

For these measurements, we have chosen $(C_7H_{10}N)_2CuBr_4$, known as DIMPY, in which $Cu^{2+}$ spins form two-leg ladders along the **a** axis of the monoclinic lattice [12–16]. The ratio $x = J_{leg}/J_{rung}$ between the leg exchange and the rung exchange is 2.2(2) in this material, placing it in the rare, strong-leg regime. The zero-field magnetic excitations measured by inelastic neutron scattering are strongly dispersive along the ladder direction but dispersionless, within the instrumental resolution, along orthogonal directions. Specific heat shows no long-range magnetic order down to 150 mK [13] demonstrating that DIMPY is an excellent 1D system with a gapless, TLL phase above the critical field $H_c = 3.0(3)$ T, at which the magnon gap vanishes.

Magnetization measurements were performed on a 13.8 μg single crystal with 67% deuteration using a micromechanical Faraday balance [17]. The device was loaded into a dilution refrigerator with the crystallographic **c** axis of the sample parallel to the magnetic field. A separate coil provided a field gradient of 2 T m$^{-1}$, which exerted the force $\vec{F} = \vec{M} \cdot \vec{\nabla}\vec{H}$ on the sample, where $\vec{M}$ is the magnetization. The device is designed to directly measure this force, with negligible contributions from magnetic torque. For a given field gradient, the output of the device is directly proportional to $M$. The proportionality constant was obtained by comparing, at 1.8 K and 4.3 K, the device outputs with the magnetization curves of a 30.5 mg single crystal measured in a SQUID magnetometer.

Figure 1(a) shows the magnetization of DIMPY as a function of magnetic field. At 300 mK the magnetization is nearly zero at low fields except for an evident contribution of paramagnetic impurities, whose concentration is about 0.9%. Upon reaching the critical field $H_c$, at which the magnon gap closes, the magnetization increases rapidly [14]. This signature of $H_c$ becomes less prominent as the temperature is raised. At 4.3 K, higher than the magnon gap $\Delta = 3.7(1)$ K [13], the magnetization curve is featureless, with a roughly constant rate of increase throughout the field range of our experiment.

Differentiating the magnetization with respect to the magnetic field yields the magnetic susceptibility as a function of the field, as shown in Fig. 1(b). The susceptibility is temperature-independent within our resolution at least up to 300 mK. In a truly 1D gapped antiferromagnet, the zero-temperature magnetic susceptibility exhibits a square-root divergence at $H_c$, a singularity characteristic of free 1D fermions to which the system can be mapped. As the temperature rises, this singularity becomes a rounded peak near $H_c$ [18]. Our low-temperature data clearly exhibit such a peak, providing strong evidence for the excellent 1D character of DIMPY in support of heat-capacity and inelastic neutron-scattering results [13].

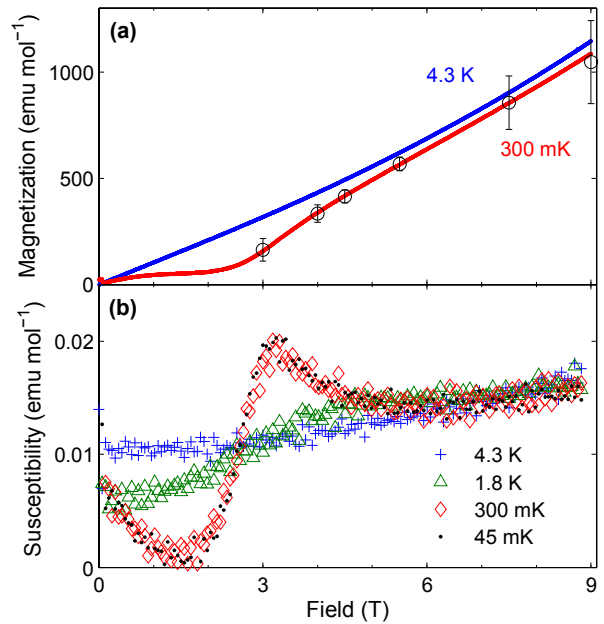

FIG. 1. (Color online). (a) Magnetization of DIMPY as a function of magnetic field at 300 mK and 4.3 K. The open circles represent DMRG results for $x = 2.0$ and $g = 2.2$. (b) Magnetic susceptibility as a function of magnetic field.

Another characteristic feature of a gapped 1D antiferromagnet is a local minimum in the magnetization at temperature $T_m$, a minimum that marks the upper limit of the TLL temperature regime [18–20]. Such minima have been observed in the $S = 1$ linear-chain compound NDMAP [21] and inferred for the $S = 1/2$ strong-rung ladder material BPCB [22], but not in any other 1D antiferromagnets because of material problems such as the presence of field-induced staggered fields and too high a critical field for the present technology. To further test the prediction of Refs. [18–20] and, at the same time, to ensure that our low-temperature susceptibility data come from the TLL regime, we have measured the magnetization as a function of temperature for different magnetic fields $H \geq H_c$, as shown in Figs. 2(a) and (b). At 4 T, slightly above $H_c$, the magnetization reaches a minimum at about 0.7 K. The minimum at higher fields occurs at even higher temperatures, assuring that our susceptibility data at 45 mK and 300 mK are indeed from deep in the TLL regime, at least at and above 4 T.

Figure 2(c) presents $T_m$ from the data as a function of

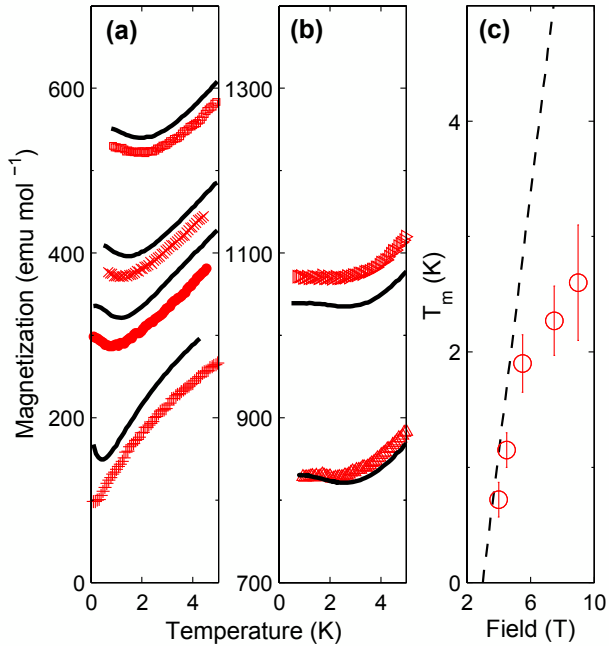

FIG. 2. (Color online). (a) Magnetization of DIMPY as a function of temperature. Symbols are experimental data, from which the magnetization of paramagnetic impurities has been subtracted; solid lines are quantum Monte Carlo simulations. Fields are, from bottom to top: 3 T, 4 T, 4.5 T, 5.5 T. (b) From bottom to top: 7.5 T, 9 T. (c) Position of the magnetization minimum from the data for $H \geq 4$ T as a function of magnetic field. Dashed line is the universal behavior, Eq. 3, for free fermions.

magnetic field, along with the universal relation

$$T_m = 0.76238 \frac{g\mu_B}{k_B}(H - H_c) \qquad (3)$$

predicted by the free fermion theory with no adjustable parameters [20]. At fields near $H_c$, where the density of fermions to which the ground-state magnons are mapped is small, the theory and experiment are in excellent agreement. As the field increases, $T_m$ falls below the universal relation, as predicted by quantum Monte Carlo (QMC) simulations [20]. This downward deviation is caused by repulsion between the fermions.

For further comparison with the experiment, we have performed QMC simulations of the magnetization of an $S = 1/2$ ladder consisting of 120 rungs, and have found that the $g$ factor $g = 2.2$ and the exchange ratio $x = 2.0$, which is consistent with 2.2(2) mentioned earlier, give the best overall agreement with the data, as shown in Figs. 2(a) and (b). At each temperature and magnetic field, 50 runs of $1.2 \times 10^7$ Monte Carlo steps were used for averaging. The simulations and the experimental data share two important features. First, the magnetization minima appear at comparable temperatures, except at 3 T which is either at, or very close to, $H_c$. Second, as the magnetic field increases, the minimum becomes shallower. Moreover, density-matrix renormalization-group (DMRG) calculations with $g = 2.2$ and $x = 2.0$—the values found by the QMC simulations—yield zero-temperature magnetization in excellent agreement with the 300 mK data, as shown in Fig. 1(a).

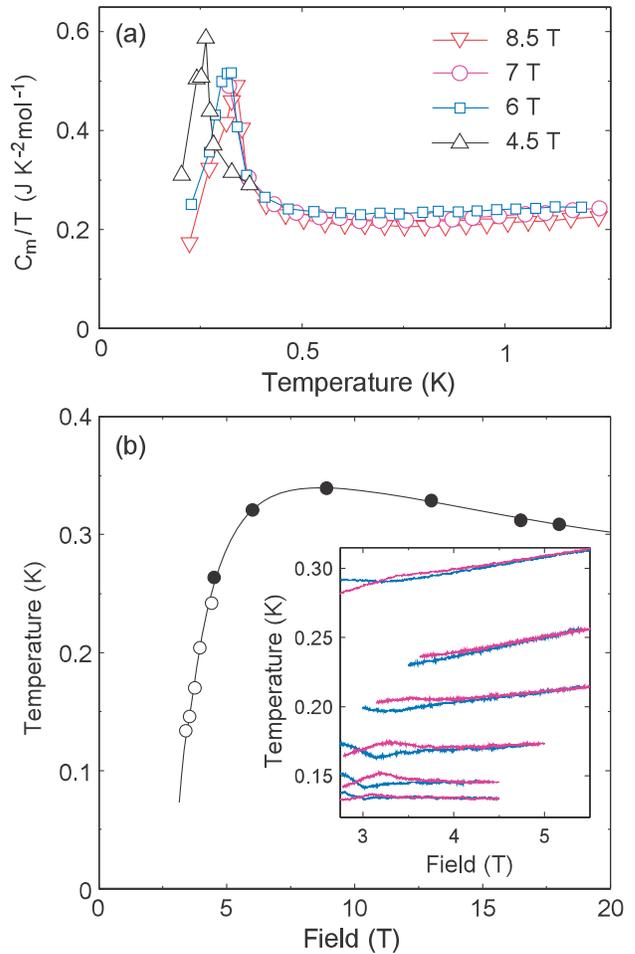

FIG. 3. (Color online). (a) Magnetic specific heat, $C_m$, of fully deuterated DIMPY, shown as $C_m/T$. (b) Ordering temperature as a function of magnetic field. Solid circles are from specific-heat peaks. Open circles are from the magnetocaloric-effect data shown in the inset, where pink (medium gray) and blue (dark gray) lines represent the sample temperature during upward and downward field sweeps, respectively. Other lines are guides to the eye.

With the excellent one-dimensionality of DIMPY firmly established, we proceed to determine the Wilson ratio $R_W$. For this purpose, we have measured the specific heat of a 6.6 mg single crystal of fully deuterated DIMPY [23] using relaxation calorimetry [24], with the magnetic field applied along the $c$ axis as in the magnetization measurements. Phonon, nuclear-spin, and nuclear-quadrupole contributions to the specific heat were subtracted as described in Ref. [13] to obtain the magnetic specific heat $C_m$. At 6 T, 7 T, and 8.9 T, $C_m$ exhibits $T$-linear behavior, demonstrated by constant $C_m/T$, at temperatures below about 1 K, except for a sharp peak signalling ordering at a lower temperature, as shown in Fig. 3(a). To detect ordering at magnetic fields close to $H_c$, we have also made magnetocaloric-effect measurements in the same calorimeter with the same sample, as shown in the inset to Fig. 3(b). From the specific-heat peaks and the magnetocaloric effect, the boundary of the ordered phase has been mapped to 18 T, as shown in Fig. 3(b). The phase boundary is highly asymmetric, peaking at about 8.9 T, where the fermion velocity is at or near maximum [13].

Ordering found in this sample at temperatures up to 340 mK is in strong contrast to the absence of ordering in a 67% deuterated sample at temperatures down to 150 mK [13], indicating that full deuteration in DIMPY dramatically enhances interladder exchanges. To further examine the effect of full deuteration, we have measured the magnetization of the sample as a function of the magnetic field at 1.8 K and 4.3 K in the SQUID magnetometer. The results were identical to those of the 67% deuterated sample within 2.0%, ensuring that the intra-ladder exchanges $J_{leg}$ and $J_{rung}$—which govern the TLL physics of DIMPY—are unaffected by full deuteration.

$C_m/T$ needed to determine $R_W$ is obtained from the data at 6 T, 7 T, and 8.9 T at temperatures below 0.75 K, excluding the region affected by the peak. At 5 T and 8 T, $C_m/T$ is taken from published data for a 67% deuterated sample [13]. For consistency, we use the susceptibility data taken at 300 mK rather than 45 mK, although the two are nearly identical. On assumption that $g=2.2$ in Eq. 1, $R_W$ determined from the susceptibility and specific heat is shown in Fig. 4 as a function of the normalized magnetization $m = M(N_A S g \mu_B)^{-1}$. It takes on values close to 4, increasing with increasing $m$.

To examine whether this result validates Eq. 2, we have employed the DMRG method to calculate the TLL parameter $K$ for $x=2.0$ [25]. The result, shown in Fig. 4 as $R_W$ by assuming Eq. 2, is in good agreement with an earlier calculation [26]. Starting from 1 at $m=0$, $K$ deviates upward from this universal free-fermion value — characteristic behavior of a strong-leg ladder [26] as opposed to a strong-rung ladder. $R_W$ from the experiment on the other hand is not always larger than 4 as expected from Eq. 2. A plausible explanation is that this quantity is sensitive to the value of the $g$ factor that enters Eq. 1. In fact, $g=2.04$ suggested by saturation-magnetization data [14] —instead of $g=2.2$ chosen here on the basis of the comparison of the magnetization data with the QMC simulations— will raise $R_W$ by about 16%, bringing it to closer agreement with $4K$ from the DMRG calculations. Taking this into account, we conclude that within the combined uncertainties of experiment and calculations, the result strongly supports Eq. 2, the equivalence between $R_W$ and $K$.

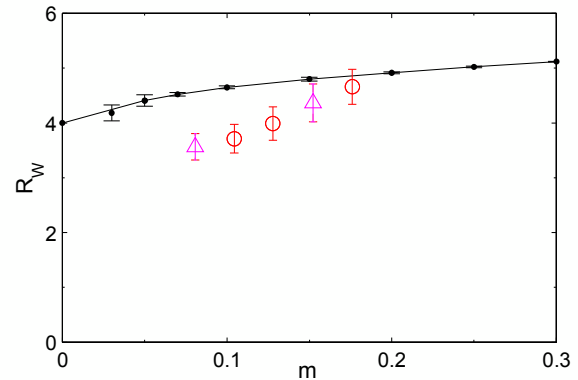

FIG. 4. (Color online). Dependence of the Wilson ratio $R_W$ of DIMPY on the normalized magnetization $m$. Open symbols are experimental data, assuming $g=2.2$ in Eq. 1; specific-heat data from Fig. 3(a) have been used for circles, and those from Ref. [13] for triangles. The error bars represent only the combined uncertainties of the magnetization and heat capacity measurements; the uncertainty of the $g$ factor has not been included. Small filled circles are the TLL parameter $K$ for $x=2.0$, computed by the DMRG method and displayed as $R_W$ by assuming Eq. 2. The line is a guide to the eye.

To summarize, our magnetization measurements further confirm that DIMPY is an ideal 1D system whose gapless phase above $H_c$ is a Tomonaga-Luttinger liquid (TLL) at low temperatures. The magnetization in this phase exhibits a minimum at $T_m$, whose dependence on magnetic field is in good agreement with quantum Monte Carlo simulations. The maximum in the low temperature susceptibility near $H_c$ is consistent with the square-root singularity expected for gapped 1D antiferromagnets at zero temperature. The successful determination of the Wilson ratio, in conjunction with the density-matrix renormalization-group calculations of the TLL parameter $K$, makes DIMPY the first laboratory 1D system that lends support to the relation $R_W = 4K$.

We thank B. Andraka and K. Ingersent for helpful discussions. Thanks are also due to J.-H. Park, T. P. Murphy, and G. E. Jones for assistance, G. W. Tremmeling for preparation of the partially deuterated sample, and M. W. Meisel for the generous loan of a magnetization standard. The calorimetric measurements were made at the National High Magnetic Field Laboratory (NHMFL), which is supported by NSF Cooperative Agreement DMR-0654119, the State of Florida, and the DOE. The QMC calculation employed the stochastic series expansion code of the ALPS project [27]. K.N. and H.B.C. were supported by the NHMFL UCGP. T.H. was partially supported by the Division of Scientific User

Facilities, Office of BES, DOE. C.H. was supported by Grant-in-Aid for Scientific Research No. 21110522 from the Ministry of Education, Culture, Sports, Science, and Technology of Japan.

---

# Supplementary information for "Wilson Ratio of a Tomonaga-Luttinger Liquid in a Spin-1/2 Heisenberg Ladder"

K. Ninios,[1,2] Tao Hong,[3] T. Manabe,[4] C. Hotta,[4] S. N. Herringer,[5] M. M. Turnbull,[5] C. P. Landee,[5] Y. Takano,[1] and H. B. Chan[2]

[1]*Department of Physics, University of Florida, Gainesville, Florida 32611, USA*
[2]*Department of Physics, The Hong Kong University of Science and Technology, Clear Water Bay, Kowloon, Hong Kong, China*
[3]*Quantum Condensed Matter Division, Oak Ridge National Laboratory, Oak Ridge, Tennessee 37831, USA*
[4]*Department of Physics, Faculty of Science, Kyoto Sangyo University, Kyoto 603-8555, Japan*
[5]*Carlson School of Chemistry and Department of Physics, Clark University, Worcester, Massachusetts 01610, USA*

The model hamiltonian for DIMPY is a two-leg Heisenberg ladder,

$$\mathcal{H} = \sum_{l=1}^{L} J_{rung} \boldsymbol{S}_{1,l} \cdot \boldsymbol{S}_{2,l} + \sum_{l=1}^{L-1} \sum_{i=1,2} J_{leg} \boldsymbol{S}_{i,l} \cdot \boldsymbol{S}_{i,l+1}, \quad (1)$$

where $\boldsymbol{S}_{i,l}$ is the $S=1/2$ spin on the $i$th leg and $l$th rung. At zero field, this model has a finite spin gap, $\Delta > 0$, for any value of $x = J_{leg}/J_{rung}$. The gapless phase above the critical field $H_c = \Delta/(g\mu_B)$ is a Tomonaga-Luttinger liquid (TLL).

To evaluate the TLL parameter $K$ of this phase with the density-matrix renormalization-group (DMRG) technique, we have adopted the procedure developed by Hikihara and Furusaki [1]. In their method, the transverse two-spin-two-spin correlation $\langle S^x_{\pi,l} S^x_{\pi,l+r}\rangle$ is computed as a function of rung position $l$ and rung-rung distance $r$, as is the longitudinal magnetization $\langle S^z_{0,l}\rangle$ as a function of $l$. Here

$$\boldsymbol{S}_{\pi,l} = \boldsymbol{S}_{1,l} - \boldsymbol{S}_{2,l}, \quad (2)$$
$$\boldsymbol{S}_{0,l} = \boldsymbol{S}_{1,l} + \boldsymbol{S}_{2,l}. \quad (3)$$

According to Ref. [1],

$$\langle S^x_{\pi,l} S^x_{\pi,l+r}\rangle = 2X(l, l+r; Q) \quad (4)$$
$$\langle S^z_{0,l}\rangle = \frac{1}{2} + z(l; Q), \quad (5)$$

where $Q = \frac{2\pi L}{L+1}(M/M_{sat} - 1/2)$ with $M_{sat}$ being the saturation magnetization and $L$ the number of rungs, and $X$ and $z$ are functions given in Ref. [1] (see Eqs. 11 and 13). Of the two, $X$ contains a number of exponents, which in turn contain $K$.

In our DMRG calculations, we kept up to 200 states per block, for a ladder comprising 100 rungs and open boundaries. The truncation error was at most on the order of $10^{-8}$ and typically on the order of $10^{-10}$. A local field, $h_{edge}$, was imposed on each boundary in order to suppress a Friedel oscillation, which arises from mixing of states that do not belong to the ground-state manifold, the so-called "boundary effect" [2]. The strength of this field was chosen to make $\langle S^z_{0,l}\rangle$ at $M/M_{sat} = 0.5$ as uniform as possible, as expected in the bulk limit. The optimum values were found to be $h_{edge}/J_{rung} = 1.275$, 1.203, 1.088, and 0.0025 for $x = 2.2$, 2.1, 2.0, and 0.01, respectively.

Subsequently, $\langle S^x_{\pi,l} S^x_{\pi,l+r}\rangle$ was computed for $l$ and $r$ that satisfied the condition $2l + r = L+1$ or $L$; in other words the two rung positions $l$ and $l+r$ were at equal, or nearly equal, distance from the ladder center. $K$ was extracted by fitting Eq. 4 to the data over 70 or 80 consecutive $r$, starting from either $r=10$ or 20, as shown in Fig. 1(a). Finally, a mean of $K$ from these four fits was taken. Figure 1(b) presents our results for the four values of $x$ given above, showing good agreement with the results of Ref. [1]. The error bars here, as well as in Fig. 4 of our paper, represent standard deviations from means.

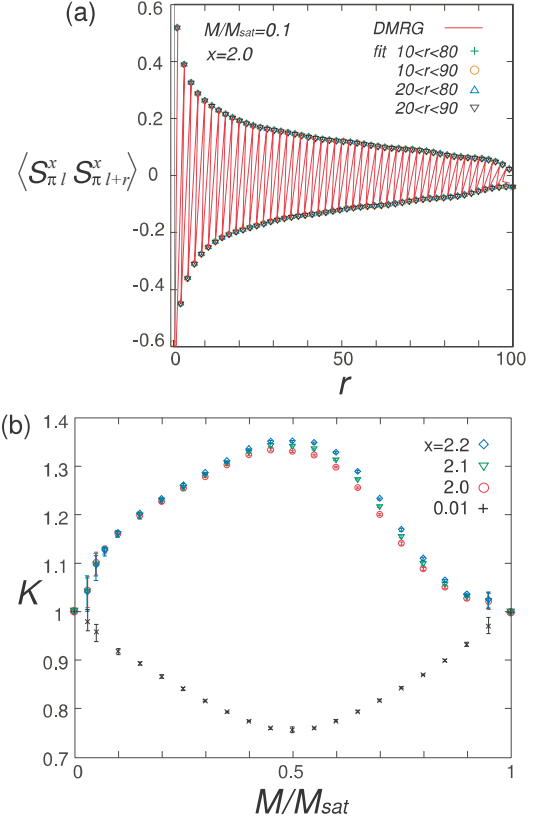

FIG. 1. (a) Transverse correlation function for four spins on two rungs, $\langle S^x_{\pi,l} S^x_{\pi,l+r}\rangle$, as a function of $r$. Line is a DMRG result for fractional magnetization $M/M_{sat} = 0.1$ and $x = J_{leg}/J_{rung} = 2.0$. Symbols indicate fits to Eq. 4 over four different fitting ranges. On this scale, the fits are indistinguishable from the data even outside those ranges. (b) TLL parameter, $K$, as a function of $M/M_{sat}$ for several values of $x$.